\documentclass[a4paper,aps,prd,10pt,twocolumn,superscriptaddress,nofootinbib,notitlepage,floatfix]{revtex4-2}

\usepackage[colorlinks=true, allcolors=blue]{hyperref}
\usepackage[total={18cm, 25.05cm}]{geometry}

\usepackage{graphicx}
\usepackage{amssymb,amsmath,amsfonts}
\usepackage{slashed}
\usepackage{bbold}
\usepackage{dcolumn}
\usepackage{bm}
\usepackage{physics}
\usepackage{mathrsfs}
\usepackage{mathtools}
\usepackage{multirow}

\newcommand \be{\begin{equation}}
\newcommand \ee{\end{equation}}
\newcommand \bea{\begin{eqnarray}}
\newcommand \eea{\end{eqnarray}}
\newcommand\m[1]{\begin{pmatrix}#1\end{pmatrix}} 

\newcommand\LFD[1]{\frac{\overrightarrow{\delta }}{\delta #1}}
\newcommand\RFD[1]{\frac{\overleftarrow{\delta }}{ \delta #1}}

\def\deltabar{{\raisebox{2.9pt}{\bm{\mbox{--}}} \mkern -9.2mu\delta}}

\newcommand{\p}{{\bf p}}
\newcommand{\q}{{\bf q}}

\begin{document}
\allowdisplaybreaks

\title{Scale dependence of the Kondo interaction in the \\functional renormalization group formalism}

\author{Gergely Fej\H{o}s}
\email{gergely.fejos@ttk.elte.hu}
\affiliation{Institute of Physics and Astronomy, E{\"o}tv{\"o}s University, 1117 Budapest, Hungary}
\author{Taro Kimura}
\email{taro.kimura@u-bourgogne.fr}
\affiliation{Institut de Math{\'e}matiques de Bourgogne, Universit{\'e} de Bourgogne, CNRS, 21078 Dijon, France}
\author{Zsolt Sz{\'e}p}
\email{szepzs@achilles.elte.hu}
\affiliation{HUN-REN-ELTE Theoretical Physics Research Group, 1117 Budapest, Hungary}
\affiliation{Institute of Physics and Astronomy, E{\"o}tv{\"o}s University, 1117 Budapest, Hungary}

\begin{abstract}
  Scale evolution of interactions between a Weyl fermion and a heavy magnetic impurity is calculated non-perturbatively using the functional renormalization group technique.
  Using an expansion around the vanishing pairing gap, we derive the flow equations for all possible quartic couplings in the system.
  We find that, contrary to conventional perturbation theory, the usual spin-spin isotropic interaction necessarily splits into two invariant parts during the scale evolution, which are fully allowed by the $SU(2)$ spin-rotation symmetry.
  We also find the existence of an infrared stable interacting fixed point, which can be responsible for intermediate-coupling screening effects.
  The calculation scheme presented here is rather general, and is expected to be easily applicable to various spin-spin-like interactions in fermionic systems.
\end{abstract} 

\maketitle

\section{Introduction}
 
It was shown by Kondo that, calculated to second order in the Born approximation, the cross section for scattering of metallic electrons by a localized magnetic impurity develops a logarithmic singularity inversely proportional to the temperature \cite{Kondo64} due to spin-flip scattering processes around the Fermi level.
As a result, the resistivity increases when the temperature is lowered, giving an explanation for the resistance minimum seen experimentally in dilute magnetic alloys \cite{Haas}.
The original model calculation done by Kondo in the case of antiferromagnetic coupling between the impurity and conduction electrons gives a diverging resistivity for vanishing temperature (Kondo problem), which indicates that standard perturbation theory is not applicable at very low temperatures.
Subsequently, it was shown \cite{K-A_rel} that, in a certain range of the parameter space, Kondo's model can be mapped onto the Anderson impurity model introduced earlier \cite{Anderson61}.

The scaling analysis of the behavior of the scattering amplitude in which the energy scale is lowered and the electronic states are integrated out at the edge of the conduction band revealed at lowest order in the coupling that there is a scale where the growing running coupling constant hits a logarithmic pole \cite{Anderson69, Anderson70}.
The corresponding dynamically generated scale is called the Kondo temperature, which turns out to be the scale at which the convergence of the Born series breaks down in Kondo's original calculation mentioned in the previous paragraph. 
Anderson's \emph{poor man's} scaling method sums up the most singular contributions of the perturbation series identified diagrammatically by Abrikosov \cite{Abrikosov}.
It can be formally generalized~\cite{Solyom-Zawa} to sum up also subleading singularities corresponding to higher order terms in the $\beta$ function.
In fact, the next to leading order correction to the $\beta$ function is of the opposite sign compared to the leading order one, indicating the presence of an infrared (IR) fixed point.
The relevance of this fixed point, obtained within a weak coupling analysis, is questionable with regards to the original Kondo problem, as the position of the fixed point is not close enough to the Gaussian one, rendering the neglected higher order contributions large.
However, there are situations, as in the weak-coupling overscreened Kondo model, where an intermediate-coupling fixed point is responsible for the low-temperature physics, rather than a strong coupling fixed point (infinite coupling limit) as is the case in the original Kondo model with exact screening \cite{Nozieres, Wilson, Nozieres75}.  
These aspects of the Kondo phenomenon were extensively studied in condensed-matter physics and became standard textbook material \cite{Hewson93, Coleman2016} (see also \cite{Nevidomsky}).
Current developments in nanotechnology and the controllable environment provided by the quantum dots led to a revival of interest in the Kondo effect \cite{revival}.

Pair breaking in superconductors due to the Kondo screening was studied theoretically with various models \cite{borkowski92, borkowski94} and experimentally in heavy fermion compounds \cite{nc_supercond}.
Nowadays, the interplay between the Kondo effect and superconductivity is studied also in topological materials like Dirac and Weyl semimetals \cite{DWS, top_supercond}, as well as in artificially engineered devices obtained by attaching a quantum dot to a superconductor, as discussed in Ref.~\cite{Zarand21}.

Although the non-Abelian nature of interaction through which electrons screen the impurity is important for the Kondo effect to take place, the spin degree of freedom is not essential, as it can be replaced by any other multivalued quantum number. The necessary ingredients for the Kondo effect are discussed for example in \cite{Yamada}. In fact, experimental evidence for spinless Kondo effect was reported to take place in quantum dot systems \cite{spinless-Kondo}. More recently, it was shown \cite{Yasui:2013xr, Hattori:2015hka} that in light quark matter containing heavy quarks as impurities the Kondo effect can be induced by the color-flipping interactions mediated by gluons. This opens the possibility for a new phase transition of the strongly interacting matter, which is currently under active investigation \cite{Yasui:2016svc, Hattori:2022lnh}.

As mentioned above, the fixed point structure of the Kondo model is questionable when it is obtained with a standard perturbative analysis, and hence one should apply non-perturbative methods to study the low-energy physics of the Kondo effect. From this point of view, there have been various studies of the Kondo problem based on non-perturbative approaches, including the numerical renormalization group method \cite{Wilson}, the exact analysis based on the Bethe ansatz \cite{Andrei, Wiegmann, Andrei:1982cr}, large $N$ analysis \cite{Coleman, Read--Newns, Parcollet--Georges}, and conformal field theory (CFT) \cite{Affleck, Affleck--Ludwig}.

In this paper, we apply the functional renormalization group (FRG) method, which represents another powerful non-perturbative tool capable of describing physical systems over the entire energy range of interest. In what follows, we introduce a calculation scheme in which one can explore in a non-perturbative fashion the low-energy behavior of the Kondo interaction between a Weyl fermion and a heavy magnetic impurity in an effective model introduced in Ref.~\cite[Sec.~III]{Kanazawa:2016ihl}. Our main goal is not only to contribute to the understanding of the renormalization group flows of the aforementioned model, but also to develop an approximate framework within the FRG that can be used for a wide range of different models that contain interactions between fermions and (generalized) impurities.

The paper is organized as follows. In Sec.~\ref{sec:model} we revisit the model introduced in \cite{Kanazawa:2016ihl}, arguing that for a consistent FRG treatment one needs to include an extended set of quartic couplings that are inevitably generated by the flow equations. The difference with respect to the perturbative scaling is also discussed. In Sec.~\ref{sec:RGflow} the flow equations for the couplings and the wave function renormalization factor of the impurity field are derived and the fixed point structure of the model is determined. We find an intermediate-coupling fixed point, where the spin rotation symmetry of the interaction is restored. We summarize in Sec.~\ref{sec:sum} and outline some potential directions of further studies. Some technical details are provided in two appendices.

\section{Effective model \label{sec:model}}

In what follows we consider a fermion-impurity model, which can be regarded as an effective theory describing the Kondo effect. Throughout the study, we consider the following ansatz for the scale-dependent effective action:
\begin{widetext}
\bea
\label{Eq:Ansatz}
\Gamma_k[\psi,\xi] &=& \int_x \bigg[ \psi^\dagger(x)\big((\partial_\tau - \mu)\mathbb{1}
  + i v_{\rm F}\, {\bm \sigma} \cdot {\bm \nabla}\big)\psi(x)
  + \frac{\Delta_k}{2}\Big(\psi^{\rm T}(x)\sigma_2\psi(x) + \psi^\dagger(x)\sigma_2\psi^*(x)\Big)
  + Z_k \xi_s^\dagger(x)\big(\partial_\tau - \mu_\xi\big)\xi_s(x)\nonumber\\
  &+& Z_k G_k \big(\psi^\dagger(x)\xi(x)\big) \big(\psi^{\rm T}(x)\xi^*(x)\big)
   + Z_k H_k \big(\psi^\dagger(x)\psi(x)\big) \big(\xi^\dagger(x)\xi(x)\big)
  + J_k \big(\psi^\dagger(x)\psi(x)\big)^2+ Z_k^2 L_k \big(\xi^\dagger(x)\xi(x)\big)^2
  \bigg], \nonumber\\
\eea
\end{widetext}
where the notation  $(AB) \equiv A_s B_s$ is used, $s$ being a spinor index. Also, $\mu$ and $\mu_\xi$ denotes chemical potentials, while $v_{\rm F}$ is the Fermi velocity, hereafter set equal to unity.  In \eqref{Eq:Ansatz} we included all the possible four-fermion interactions between the Weyl fermion $\psi$ and the impurity $\xi$, as well as their self-interactions. It turns out that (\ref{Eq:Ansatz}) contains the minimal set of couplings necessary to be introduced so that a closed system of equations for their respective renormalization group flows can be derived.

We would like to stress that, although we have two terms describing the interaction between $\psi$ and $\xi$ fields, the couplings $G_k$ and $H_k$ are not related to the couplings $J_\perp(\equiv J_x=J_y)$ and $J_z$ of the anisotropic Kondo model (see, e.g., Ref.~\cite[p.~648]{Coleman2016}), as our model is spin-spin isotropic. Note that the usual interaction constructed from spin operators that is invariant under global $SU(2)$ spin-rotation symmetry has the form 
\be
\label{Eq:isotrop-Kondo}
\mathcal{J} \vec{S}_{\psi} \cdot \vec{S}_{\xi},
\vspace*{0.3cm}
\ee
with $\vec{S}_{\psi}=\psi^*_i \vec{\sigma}_{ij} \psi_j$ and $\vec{S}_{\xi}=\xi^*_i \vec{\sigma}_{ij} \xi_j$. This is the interaction of the isotropic Kondo model. Using the identity
\be
\label{Eq:identity}
\vec{\sigma}_{ij}\cdot \vec{\sigma}_{kl} = 2\delta_{il}\delta_{jk}-\delta_{ij}\delta_{kl}
\ee
satisfied by the Pauli matrices, one sees that the two $SU(2)$ invariants $\big(\psi^\dagger\xi\big) \big(\psi^{\rm T}\xi^*\big)$ and $\big(\psi^\dagger\psi\big) \big(\xi^\dagger\xi\big)$ appear with a fixed ratio of their coefficients.  {\it A priori} it cannot be assumed that this relation between the coefficients is preserved throughout the non-perturbative RG flow, as opposed to its perturbative counterpart, to be reviewed at the end of this subsection.\footnote[\numexpr \value{footnote}+1\relax]{A somewhat similar splitting occurs in the renormalization of the two-particle irreducible formalism, where counterterms must be introduced to a larger number of invariants of a given symmetry group than in the case of a perturbative renormalization \cite{Berges:2005hc,Fejos:2007ec}.} Based on similar grounds, purely $\psi$ and $\xi$-type quartic interactions also need to be introduced into the effective action.

The effective model defined by \eqref{Eq:Ansatz} can be regarded as an extension of that introduced in Ref.~\cite{Kanazawa:2016ihl}, where only the coupling $G_k$ and the wave function renormalization factor $Z_k$ for the $\xi$ field were used in a perturbative study. Here we have three additional couplings (i.e., $H_k$, $J_k$, $L_k$), which, as explained in the previous paragraph, cannot be dropped from the description as radiative corrections inevitably generate them. Note that we have not introduced any wave function renormalization for the $\psi$ field. We will come back to this point at the end of the next section.

Following the procedure of Ref.~\cite{Mesterhazy:2012ei} for the formulation of a fermionic FRG approach, the fields and their complex conjugates are collected in the variable $\Phi$. The kinetic part of the ansatz \eqref{Eq:Ansatz} can be written in momentum space as follows [$P=(i p_0,\p)$]:
\begin{alignat}{1}
  \label{Eq:kinetic_matrix}
  &\!\!\!\!\!\!\int_P\Phi^{\rm T}(-P) \mathcal{D}_0^{-1}(P,-P)\Phi(P) \nonumber\\
  =&\int_P \m{\psi^{\rm T}(-P),&\psi^\dagger(P),&\xi^{\rm T}(-P),&\xi^\dagger(P)}\nonumber\\
   &\times \m{0 & -\hat D^{-\rm T}(-P)  & 0 & 0 \\
    \hat D^{-1}(P) & 0 & 0 & 0\\
    0 & 0 & 0 & -Z_k\hat d^{-\rm T}(-P) \\
    0 & 0 & Z_k\hat d^{-1}(P) & 0
  } \nonumber\\
&\times \m{\psi(P)\\ \psi^*(-P)\\ \xi(P)\\ \xi^*(-P)},
\end{alignat}
with the notations $\displaystyle \int_P=\int_{-\infty}^{\infty} d p_0 \int \frac{{\rm d}^3p}{(2\pi)^3}$ and $\hat D^{-\rm T}\equiv \big(\hat D^{-1}\big)^{\rm T}$, and the $2\times 2$ inverse propagator matrices 
\bea
\hat D^{-1}(P)&=&-i p_0-\mu - \p\cdot\bm{\sigma},\nonumber\\
\hat d^{-1}(P)&=&(-i p_0 - \mu_\xi)\mathbb{1}.
\eea
In writing \eqref{Eq:kinetic_matrix} we used the fact that the matrix obtained when taking the transposed of the kinetic term for the $\psi$ field can be written as\footnote{Since the propagator of the $\xi$ field is proportional to the unit matrix, keeping the transposed notation is useful only when doing formal manipulations.}
\bea
 -i p_0+\mu - \p\cdot\bm{\sigma}^{\rm T}&=&-[i p_0-\mu + \p\cdot\bm{\sigma}]^{\rm T} \nonumber\\
 &=& -\big[(-i p_0-\mu - \p\cdot\bm{\sigma})\big|_{P\to -P} \big]^{\rm T}\nonumber\\
 &=& -\hat D^{-\rm T}(-P).
 \eea
 Note that we found it convenient not to include the wave function renormalization factor in the $2\times 2$ propagator matrix of the $\xi$ field.

The regulator matrix is chosen to have the property of the tree-level propagator matrix, namely:
\be
\label{Eq:reg_matrix}
\hspace{0.0cm}R_k(P,-Q) = \deltabar(P-Q)\m{0& \hat R^{\rm T}_k(-P) & 0 & 0 \\
  -\hat R_k(P) & 0 & 0 & 0\\
  0 & 0 & 0 & 0 \\
  0 & 0 & 0 & 0}, 
\ee
where $\deltabar(P-Q)=(2\pi)^4\delta(p_0-q_0)\delta(\p-\q)$. In our formulation only the three-momentum is regularized, thus (\ref{Eq:reg_matrix}) has nonzero components only in the $\psi$ sector. Using the notations $p\equiv|\p|$ and $\hat{\p}=\p/p$, the 2$\times$2 regulator matrix is
\bea
\label{Eq:reg_fv}
\hat R_k(P) &=& \hat\p\cdot\bm{\sigma}\, \mathcal{R}_k(P),\nonumber\\
\mathcal{R}_k(P)&=&\big[\textrm{sgn}(p-\mu)k-(p-\mu)\big]\Theta\big(k^2-(p-\mu)^2\big)\,,\quad
\eea
with the regulator function $\mathcal{R}_k(P)$ adopted from Ref.~\cite{Diehl:2009ma}, where it was used in the study of the BCS-BEC crossover.

The regulated tree-level inverse propagator matrix is defined as $\mathcal{D}_{R,k}^{-1} := \mathcal{D}_0^{-1} + R_k$, whose inverse is
\bea
&&\mathcal{D}_{R,k}(P,-Q) = \deltabar(P-Q)\\
&&\times\m{0 & \hat D_{R,k}(P) & 0 & 0 \\
  -\hat D_{R,k}^{\rm T}(-P)  & 0 & 0 & 0\\
  0 & 0 & 0 & Z_k^{-1}\hat d(P) \\
  0 & 0 & -Z_k^{-1}\hat d^{\rm T}(-P) & 0
}\nonumber
\eea
where $\p_R$ is the regulated momentum and $\hat D_{R,k}(P)$ and $\hat d(P)$ are $2\times 2$ propagator matrices:
\bea
\label{Eq:reg_pert_prop}
\hat D_{R,k}(P) = \frac{1}{-i p_0-\mu - \p_R\cdot\bm{\sigma}},\nonumber\\
\hat d(P)=\frac{\mathbb{1}}{-i p_0 - \mu_\xi}=:d(P)\mathbb{1}.
\eea
The components of the regulated three-momentum are
\be
p_{R,i} = p_i + \hat p_i\,\mathcal{R}_k(P)
= \left\{
\begin{array}{cl}
  (\mu - k)\hat p_i\  & \text{if } \mu-k\le p <\mu,\\[5pt]
  (\mu + k)\hat p_i\  & \text{if } \mu < p \le \mu + k,\\[5pt]
  p_i\  & \text{otherwise},
\end{array}
\right.
\ee
hence the propagations of the $\psi$ modes close to both sides of the Fermi surface are suppressed in a symmetrical fashion. We also have
\bea
\!\!\!\!\!\!\!\partial_k \mathcal{R}_k(P) &=& \textrm{sgn}(p-\mu) \Theta\big(k^2-(p-\mu)^2\big) \nonumber\\
&=& \Theta(\mu\le p \le \mu + k) -  \Theta(\mu-k\le p \le \mu)\,, \quad
\eea
as the Dirac delta generated from the theta function upon derivation does not contribute, as it gives
\bea
(\mu + k - p) \delta(p-(k+\mu)) \qquad \textrm{for} \qquad p\ge \mu + k \nonumber
\eea
and
\bea
(\mu - k - p) \delta(p-(\mu - k)) \qquad \textrm{for} \qquad p\ge \mu - k, \nonumber
\eea
and both terms vanish under a $p$-integral. This represents a nice feature of the chosen regulator function.

\begin{figure}[t!]
\begin{center}
\includegraphics[width=0.475\textwidth]{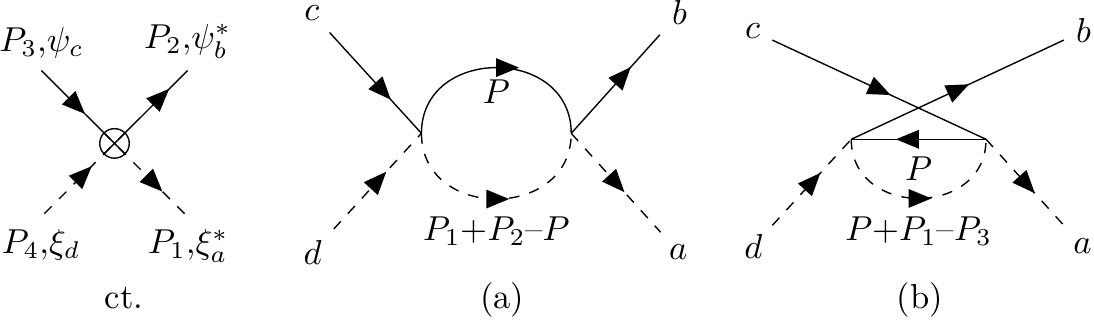}
\caption{Counterterm (ct.) absorbing the cutoff (D) dependence of the graphs (a) and (b), calculated with the vertex $-\mathcal{J}\vec\sigma_{bc}\cdot\vec\sigma_{ad}$. Internal lines are 2$\times$2 matrix propagators. We take $P_2=P_3=0$ and $P_1=P_4\equiv Q=(-i\omega,\bf 0)$.
  \label{Fig:pert_scaling} }
\end{center}
\end{figure}

Before presenting the FRG calculation,  using the formalism of Ref.~\cite{Kanazawa:2016ihl} we review the well-known fact that fluctuations with energy very close to the Fermi surface maintain the isotropic form of the spin-spin interaction \eqref{Eq:isotrop-Kondo}. The contributions to be taken into account are those used in the poor man’s scaling approach. They are depicted in Fig.~\ref{Fig:pert_scaling}. Rather than investigating how the effective coupling changes by integrating out fluctuations in a momentum shell, we perform our calculation by excluding the fluctuations around the Fermi surface, i.e., for $\mu-D\le p\le\mu+D$ with $D\ll \mu$, and apply a counterterm to absorb the cutoff dependence. A straightforward application of the Feynman rules gives
\begin{subequations}
  \label{Eq:graphs_ab}
  \bea
  \textrm{graph (a)}&=&-\mathcal{J}^2 \big(\sigma^m \sigma^n\big)_{bc} \big(\sigma^m \sigma^n\big)_{ad} I^D_2(Q) \, , \quad\\
  \textrm{graph (b)}&=&-\mathcal{J}^2 \big(\sigma^m \sigma^n\big)_{bc} \big(\sigma^n \sigma^m\big)_{ad} I^D_1(Q)\, ,\quad
  \eea
\end{subequations}
with the integrals 
\begin{subequations}
\label{Eq:pert_I1_I2}
  \bea
  I^D_1(Q)&=& \int_P^D \frac{i(p_0+\omega) + \mu}{\big(i(p_0+\omega)+\mu\big)^2-p^2}\,\frac{1}{i p_0+\mu_\xi}\,, \qquad
  \label{Eq:pert_I1}\\
  I^D_2(Q) &=& \int_P^D \frac{\mu-i(p_0+\omega)}{\big(i(p_0+\omega)-\mu\big)^2-p^2}\, \frac{1}{i p_0+\mu_\xi}\,,\qquad
  \eea
\end{subequations}
where the superscript indicates that $p\notin[\mu-D,\mu+D]$. The relation of these integrals to those appearing in the particle–particle and particle–hole channels of the non-relativistic self-energy can be given based on Eqs.~(A4) and (A5) of Ref.~\cite{Kanazawa:2016ihl}. 

In Eq.~\eqref{Eq:graphs_ab} the Pauli matrices within the first and second pairs of round brackets are associated with $\psi$ and $\xi$ fields, respectively. Note, that the order of Pauli matrices within a pair of round brackets is reversed in case of the crossed graph [graph (b)] compared to graph (a). This generates the commutator of two Pauli matrices in the sum of the singular contributions of the two graphs, as the singular part of the integrals evaluated in Appendix~\ref{app:int} is opposite in sign:
\be
\label{Eq:pert_I1_I2_sing}
\lim_{i\omega-\mu_\xi\to 0} I^D_i(-i\omega,\bm{0}) = (-1)^{i+1}\frac{\mu^2}{4\pi^2}\ln\frac{D}{\mu} + \mathcal{O}(D^0).
\ee
Therefore, the singular contribution of the two graphs gives, upon using the identity $[\sigma^m,\sigma^n]=2 i \epsilon^{mnk}\sigma^k$,
\be
\big[\textrm{graph (a)} + \textrm{graph (b)}\big]_\textrm{sing} = 2 \mathcal{J}^2 \vec\sigma_{bc}\cdot \vec\sigma_{ad} \Big(-\frac{\mu^2}{2\pi^2}\ln\frac{D}{\mu}\Big),
\ee
which has the form of the interaction in \eqref{Eq:isotrop-Kondo}, and hence can be absorbed by a counterterm: $\delta_{\bar{\mathcal J}} + 2 {\bar{\mathcal J}}^2\ln\frac{D}{\mu}=0$, where we defined $\bar{\mathcal J}:=\frac{\mu^2}{2\pi^2}\mathcal{J}$.

This procedure gives the same one-loop $\beta$ function as the poor man's scaling:
\be
\beta^{(1)}_{\bar{\mathcal J}}=D \frac{\partial}{\partial D}\delta_{\bar {\mathcal J}}=-2{\bar {\mathcal J}}^2.
\ee

\section{Renormalization group equations \label{sec:RGflow}}

\subsection{Coupling flows}
The evolution equation for the quantum effective action reads (STr denotes supertrace)
\bea
\label{Eq:WE_compact}
\partial_k \Gamma_k[\Phi] &=& \frac{1}{2}\textrm{STr}\Big[\partial_k R_k \, \big(\Gamma_k^{(1,1)}[\Phi] + R_k\big)^{-1}\Big]
  \nonumber\\
  &=&  \frac{1}{2}\textrm{STr}\, \tilde\partial_k \ln \big(\Gamma_k^{(1,1)}[\Phi] + R_k\big),
\eea
where
\[
\big(\Gamma_k^{(1,1)}[\Phi]\big)_{ij}(P,-Q) := \LFD{\Phi^{\rm T}_i(-P)}\Gamma_k[\Phi] \RFD{\Phi_j(Q)},
\]
with $i$ and $j$ labeling components of the variables $\Phi$ and $\Phi^{\rm T}$ introduced in Eq.~\eqref{Eq:kinetic_matrix}. In (\ref{Eq:WE_compact}) $\tilde\partial_k$ is defined to act only on the $k$ dependence of the regulator $R_k$.

Writing the full scale-dependent inverse propagator as 
\be
\label{Eq:Dyson}
\mathcal{G}^{-1}_{R,k}[\Phi]:=\Gamma_k^{(1,1)}[\Phi] + R_k = \mathcal{D}_{R,k}^{-1} + \Sigma_k[\Phi],
\ee
where $\Sigma_k[\Phi]:=\Gamma_k^{(1,1)}[\Phi] - \mathcal{D}_0^{-1}$, the functional series expansion of Eq.~\eqref{Eq:WE_compact} gives
\bea
\label{Eq:WE_field_exp}
\partial_k \Gamma_k[\Phi] &=& \frac{1}{2}\textrm{STr}\, \tilde\partial_k\ln \mathcal{D}_{R,k}^{-1}\nonumber\\
 &&+  \frac{1}{2}\textrm{STr}\,\tilde\partial_k \sum_{n=1}\frac{(-1)^{n+1}}{n} \big[\mathcal{D}_{R,k} \, \Sigma_k[\Phi]\big]^n\,.\qquad
\eea
Taking into account the extra minus sign contained in the definition of the supertrace, the second term ($n=2$) in this sum has in momentum space the following form [$\mathcal{D}_{R,k}(P)\equiv \mathcal{D}_{R,k}(P,-P)$]
\bea
\label{Eq:Mercator_term2}
&&{\rm RHS}_{n=2} \equiv \frac{1}{4}\int_{P}\int_{Q}\tilde\partial_k \tr\big[ \mathcal{D}_{R,k}(P)\,\Sigma_k[\Phi](P,-Q)\nonumber\\
&&\hspace{3.5cm}\times\mathcal{D}_{R,k}(Q)\,\Sigma_k[\Phi](Q,-P) \big].\qquad
\eea
In each element of the $4\times4$ block matrix $\Sigma_k[\Phi]$, the ansatz \eqref{Eq:Ansatz} generates terms that contain the product of two fermionic fields. These combine in (\ref{Eq:Mercator_term2}) into products of four fermionic fields contributing to the right-hand side (RHS) of the flow equation (\ref{Eq:WE_compact}). The terms in question have to be matched against similar contributions given by the ansatz on the left-hand side of \eqref{Eq:WE_compact}. These projections onto specific four-point interactions of the ansatz lead to the flow equation of the chosen coupling. Note that the 11 and 22 matrix elements of $\Sigma_k[\Phi]$ also contain a field-independent contribution proportional to the coupling $\Delta_k$. As a result of this, each $n\ge3$ term of the expansion in \eqref{Eq:WE_field_exp} also generates a term containing the product of exactly four fields multiplied with some nonzero power of $\Delta_k$. [In other words, the second term of the expansion, that is \eqref{Eq:Mercator_term2}, only gives the leading order flow equations in the expansion in $\Delta_k$.] Note, however, that in the flow of $\Delta_k$ itself (obtained at the order of two fields in the expansion) at least one $\Delta_k$ factor appears in each term of the right-hand side of (\ref{Eq:WE_field_exp}). That is, $\Delta_k = 0$ is always a fixed point of the corresponding flow equation, whose stability will be investigated in Sec.~\ref{SS:FP}. From here onward, we drop all the $\Delta_k$ dependence and evaluate the remaining scale evolutions in the $\Delta_k = 0$ fixed point. That is to say, in the vicinity of this fixed point the $n=2$ term in (\ref{Eq:WE_field_exp}) provides the exact flows within our ansatz for the effective action. That is also to say, there is no small parameter in terms of which the expansion is realized, showing the non-perturbative nature of the approach.

\begin{figure*}[t!]
\begin{center}
\includegraphics[width=0.8\textwidth]{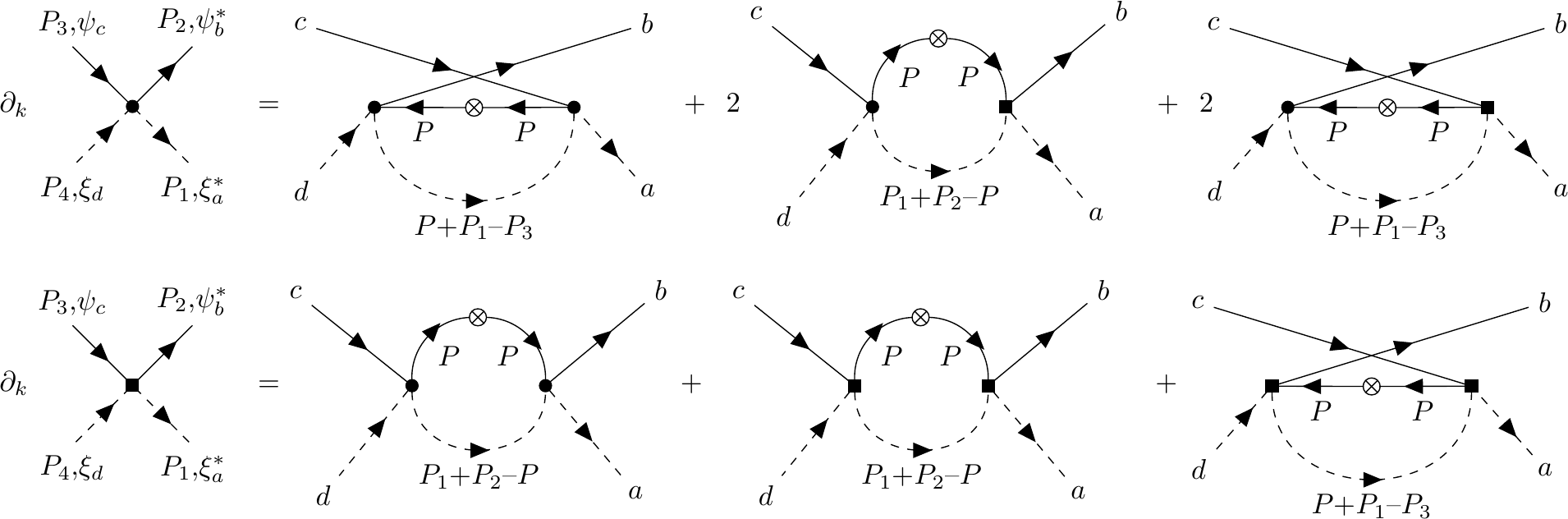}
\caption{Graphical representation of the relevant part [see text above \eqref{Eq:G_H_flows}] of the flow equation for the couplings $G_k$ (first row) and $H_k$ (second row). Internal lines represent $2\times 2$ matrix propagators, the crossed dots represent $\partial_k \hat R_k$, while the external legs indicate fields. The Feynman rule for the vertex represented with a \emph{blob} is $-Z_k G_k\delta_{ac}\delta_{bd}$, while for the one represented with a \emph{square} is $-Z_k H_k\delta_{ad}\delta_{bc}$, as obtained from the ansatz \eqref{Eq:Ansatz} by taking the field derivatives in the order indicated in Eq.~\eqref{Eq:Gamma22_def}. The momenta flow in the direction of the arrows and for all graphs the labeling of the external momenta is the same as in the case of the vertex appearing on the left-hand sides of the equations. \label{Fig:flow-graphs} }
\end{center}
\end{figure*}

Since according to Ref.~\cite{Kanazawa:2016ihl} the flow equations for the couplings are to be considered with nonzero frequency associated to the impurity fields, we have to work with an inhomogeneous background when calculating $\Gamma_k^{(1,1)}[\Phi]$. This leads to a momentum-dependent $4\times4$ block matrix $\Sigma^{ab}_k[\Phi]$ with elements depending on the two spinor indices $a$ and $b$. These matrix elements are given in Appendix~\ref{app:Sigma_elem}.

The complete set of flow equations obtained from \eqref{Eq:Mercator_term2} with the method sketched above is given in Appendix~\ref{app:int}. 
It turns out that the flows of $J_k$ and $Z_k^2 L_k$ have no influence on the flows of $Z_k G_k$ and $Z_k H_k$ in the vicinity of the $\Delta_k\equiv 0$ fixed point (see the end of Sec. ~\ref{SS:FP}).  The latter are given by 
\begin{subequations}
\label{Eq:G_H_flows}  
\bea
\label{Eq:G-flow}
\partial_k (Z_k G_k) &=& -2 Z_k G_k^2 I_1 - 2 Z_k G_k H_k (I_1+I_2),\quad\\
\label{Eq:H-flow}
\partial_k (Z_k H_k) &=& -Z_k G_k^2 I_2 - Z_k H_k^2 (I_1+I_2),\quad 
\eea
\end{subequations}
where $\displaystyle I_i= \lim_{i\omega-\mu_\xi\to 0} I_i(-i\omega,\bm{0})$ stands for the limits of the following integrals
\begin{subequations}
\bea
I_1(Q)&=&\frac{1}{2} \int_P \tilde\partial_k \tr\big[\hat D_{R,k}(P)\hat d(P+Q) \big] \nonumber\\
&=& \frac{1}{4} \int_P \tilde\partial_k \tr\big[\hat D_{R,k}(P)\big]\,\tr\big[\hat d(P+Q)\big]\nonumber\\
&=& \int_P \tilde \partial_k
\frac{i(p_0+\omega) + \mu}{\big(i(p_0+\omega)+\mu\big)^2 - p_R^2}
\,\frac{1}{i p_0+\mu_\xi}\,, \qquad
\label{Eq:I1}\\
I_2(Q)&=&\frac{1}{4} \int_P \tilde\partial_k \tr\big[\hat D_{R,k}(P)\big]\tr\big[\hat d(Q-P) \big]\nonumber\\
&=& \int_P \tilde\partial_k
\frac{\mu-i(p_0+\omega)}{\big(i(p_0+\omega)-\mu\big)^2 - p_R^2}\, 
\frac{1}{i p_0+\mu_\xi}\,.\qquad
\eea
\end{subequations}

A graphical illustration of the flow equations \eqref{Eq:G_H_flows} is given in Fig.~\ref{Fig:flow-graphs} based on the fact that diagrams contribute to proper vertices with an additional minus sign. The graphs in Fig.~\ref{Fig:flow-graphs} are built up in terms of the $2\times 2$ propagator matrices, which can be constructed based on \eqref{Eq:Ansatz} and the regulator matrix. The factor of 2 in front of the two graphs in the first row of Fig.~\ref{Fig:flow-graphs} reflects the fact that a graph in which the position of the vertices is interchanged compared to the one drawn gives an identical contribution. 

For a fixed point analysis, one needs to take the $k\rightarrow 0$ limit. Calculating the integrals for $k\ll \mu$, one obtains
\be
\label{Eq:I1_2_res}
I_1= \frac{\mu^2}{4\pi^2\,k} + \mathcal{O}(k^0) \quad \text{and} \quad
I_2=-\frac{\mu^2}{4\pi^2\,k} + \mathcal{O}(k^0).
\ee
Using the above results we see that the second term on the right-hand side of Eqs.~\eqref{Eq:G-flow} and \eqref{Eq:H-flow} does not play any role in the fixed point analysis, since when the flow equation is multiplied by $k$ (i.e., taking the logarithmic derivative) its singular part cancels. Concerning the regular part, the $\mathcal{O}(k^0)$ term in the expansion of $I_1$ and $I_2$ differs, thus $I_1+I_2\ne 0$ at a general scale $k$. 

We emphasize that the obtained flow equations are incompatible with the findings of perturbation theory in the sense that if one wants to prescribe the ratio $G_k/H_k \equiv -2$ as it is done in case of the isotropic spin-spin interaction, then (\ref{Eq:G-flow}) and (\ref{Eq:H-flow}) contradict each other  at a general scale $k$. That is, as opposed to the perturbative RG (see the end of Sec. \ref{sec:model}), the $G$ and $H$ couplings do scale differently according to the non-perturbative renormalization group flows. Only in the deep infrared $I_1+I_2\to 0$ and the evolution is described by a single $\beta$ function, as in the case of the perturbative RG.

\subsection{Wave function renormalization flows}
For the scale evolution of the wave function renormalization function one has to consider the second derivative of the flow equation that removes the impurity fields in the quadratic part of the ansatz \eqref{Eq:Ansatz}. From \eqref{Eq:WE_compact} one obtains with the notation introduced in \eqref{Eq:Dyson},
\begin{align}
\label{Eq:dkZk_general}
&\LFD{\xi^\dagger}\partial_k\Gamma_k[\Phi] \RFD{\xi} =  \frac{1}{2}\textrm{STr}\, \tilde\partial_k\,  \LFD{\xi^\dagger} \ln \mathcal{G}_{R,k}^{-1}[\Phi]\RFD{\xi} \nonumber\\
&=  -\frac{1}{2}\textrm{STr}\bigg[ \partial_k R_k \, \mathcal{G}_{R,k}[\Phi] \,\bigg(\LFD{\xi^\dagger}\Gamma_k^{(1,1)}[\Phi]\RFD{\xi} \bigg)  \, \mathcal{G}_{R,k}[\Phi]\bigg],
\end{align}
where we used the fact that there are no three-point effective vertices and, for simplicity, we have not indicated the spinor indices and the momenta of $\xi^{\dagger}$ and $\xi$.

In order to obtain the flow equation for $Z_k$, one is allowed to replace the regularized full propagator $\mathcal{G}_{R,k}[\Phi]$ with its tree-level counterpart, $\mathcal{D}_{R,k}$. The difference between the two gives higher order contributions in terms of the field variables, therefore, it does not contribute to the flow of $Z_k$. It is easy to see that due to the structure of the regulator the trace gets contribution only from the $\psi$ sector. A factor of 2 comes from the $4\times 4$ block structure due to the property
\be
\label{Eq:Gamma22_property}
\big(\Gamma^{(1,1)}_{k,\psi^\dagger\psi}[\Phi](P,-Q))^{\rm T}=-\Gamma^{(1,1)}_{k,\psi^{\rm T}\psi^*}[\Phi](-Q,P), 
\ee
which is manifest on the matrix elements given in the second and third rows of \eqref{Eq:Sigma_elements}. Thus, in terms of $2\times 2$ matrices, one obtains from \eqref{Eq:dkZk_general} 
\bea
\label{Eq:dkZk_LO}
\hspace*{-0.35cm}&&\LFD{\xi^\dagger}\partial_k\Gamma_k[\Phi] \RFD{\xi}\bigg|_{\Phi=0}=  -\int_P \tr \bigg[\partial_k\hat R_k(P)\, \hat D_{R,k}(P) \, \nonumber\\
  \hspace*{-0.35cm}&&\hspace{1cm}\times \LFD{\xi^\dagger} \Gamma_{k,\psi^\dagger\psi}^{(1,1)}[\Phi](P,-P) \RFD{\xi}\, \hat D_{R,k}(P)\bigg].
\eea
Note that the minus sign here comes from the 21 element of the $4\times 4$ regulator block matrix \eqref{Eq:reg_matrix}, as the minus sign from the supertrace is already taken into account. 

\begin{figure}[!th]
\centering
\includegraphics[width=0.35\textwidth]{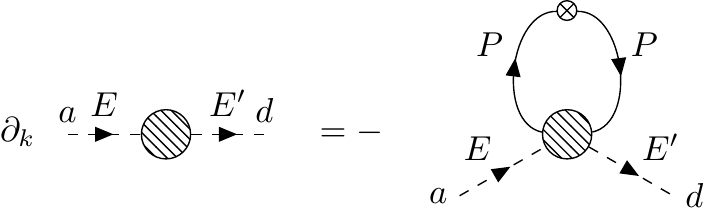}
\caption{Flow equation of the wave function renormalization factor. The blobs represent $\Gamma_k^{(1,1)}$ on the left-hand side and  $\Gamma_k^{(2,2)}$ on the right-hand side of the equation.\label{Fig:dkZk}}
\end{figure}

Using the ansatz on the left-hand side of \eqref{Eq:dkZk_LO}, one obtains the flow equation
\bea
\label{Eq:dkZk-LO}
&&\deltabar(E-E')\delta_{a d} (E_0+\mu_\xi)\partial_k Z_k\nonumber\\
&&=\int_P\partial_k \mathcal{R}_k(P)\,[\hat\p\cdot \bm{\sigma}]_{s s'}\, [\hat D_{R,k}(P)]_{s'b}\, \nonumber\\
&&\hspace{0.5cm}\times \Gamma^{(2,2)}_{k,\xi^*_a \psi_b^* \psi_c \xi_d}[\Phi](E,P,-P,-E')\, [\hat D_{R,k}(P)]_{cs}, \quad
\eea
where the effective four-point vertex
\bea
&&\Gamma^{(2,2)}_{k,\xi^*_a \psi_b^* \psi_c \xi_d}[\Phi](P_1,P_2,-P_3,-P_4)\nonumber\\
&&:=\LFD{\xi_a^*(P_1)}\LFD{\psi_b^*(P_2)}\Gamma_k[\Phi]\RFD{\psi_c(P_3)}\RFD{\xi_d(P_4)}\nonumber\\
&&=\LFD{\xi^*_a(P_1)} \Gamma_{k,\psi_b^*\psi_c}^{(1,1)}[\Phi](P_2,-P_3) \RFD{\xi_d(P_4)},\qquad
\label{Eq:Gamma22_def}
\eea
is defined with the convention that on both sides the derivative nearest to the Grassmann functional $\Gamma_k[\Phi]$ acts first. Note that Eq.~\eqref{Eq:dkZk-LO} can be also written using Fig.~\ref{Fig:dkZk} by replacing the two-point function $\Gamma_{k,\xi^\dagger\xi}^{(1,1)}$ with $Z_k d^{-1}$. 

In what follows, in order to obtain the flow of the wave function renormalization factor, we use in \eqref {Eq:dkZk-LO} the \emph{one-loop approximation} of $\Gamma_{k,\xi^*\psi^*\psi\xi}^{(2,2)}$. This represents the simplest non-trivial choice, and should be considered as part of the approximate solution presented here. In fact, the one-loop expression of $\Gamma_{k,\xi^*\psi^*\psi\xi}^{(2,2)}$ is already known, as it can be derived from \eqref{Eq:Mercator_term2} by approximating the operation $\tilde\partial_k \rightarrow \partial_k$, and integrating over $k$. Alternatively, by recalling that in perturbation theory the one-loop expression of the effective action is given by (see, e.g., Sec.~IV of \cite{Jaeckel:2002rm})
\be
\label{Eq:Gamma_k-1loop}
\Gamma_k^\text{1-loop}[\Phi] =  \frac{1}{2}\textrm{STr}\,\ln \big(\Gamma_k^{(1,1)}[\Phi] + R_k\big),
\ee
the one-loop four-point vertex can be obtained by using the decomposition introduced in Eq.~\eqref{Eq:Dyson} and keeping the second term in the functional series expansion \eqref{Eq:WE_field_exp}.

The one-loop expression of the four-point vertex $\Gamma^{(2,2)}_{k,\xi^*_a \psi_b^* \psi_c \xi_d}$ is given in agreement with Ref.~\cite{Kanazawa:2016ihl} by a bubble integral containing two different kinds of propagators, one of the impurity and one of the $\psi$ field. Pictorially its expression can be seen by removing the insertion (crossed dot) representing $\partial_k \hat R_k$ from the one-loop diagrams given in Fig.~\ref{Fig:flow-graphs}. This leads to a single propagator line of the $\psi$ field. Note that we have $Z_k^2$ from the vertices and a factor of $Z_k^{-1}$ from the impurity propagator, so an overall factor of $Z_k$ in front ($\hat D\equiv \hat D_{R,k}$):
\begin{widetext}
\bea
&&\Gamma^{(2,2)}_{k,\xi^*_a \psi_b^* \psi_c \xi_d}[\Phi](P_1,P_2,-P_3,-P_4) \nonumber\\
&&\hspace{1cm}= -Z_k\, \deltabar(P_1+P_2-P_3-P_4) \bigg\{G_k^2 \bigg[\delta_{bd}\delta_{ac}\int_Q \tr\big[\hat D(Q) \hat d(P_1-P_3+Q)\big]+ \int_Q \hat D_{ad}(Q) \hat d_{bc}(P_1+P_2-Q)\bigg]\nonumber\\
&&\hspace{5.4cm}+H_k^2 \bigg[\int_Q \hat D_{bc}(Q) \hat d_{ad}(P_1-P_3+Q) +
  \int_Q \hat D_{bc}(Q) \hat d_{ad}(P_1+P_2-Q) \bigg]\nonumber\\
&&\hspace{5.4cm}+G_k H_k \bigg[
\delta_{ac} \int_Q \hat D_{bb'}(Q) \hat d_{b'd}(P_1-P_3+Q)
+ \delta_{bd} \int_Q \hat d_{ab'}(P_1-P_3+Q) \hat d_{b'c}(Q)\nonumber\\
&&\hspace{6.8cm}+\int_Q \hat D_{bd}(Q) \hat d_{ac}(P_1+P_2-Q)
+\int_Q \hat D_{ac}(Q) \hat d_{bd}(P_1+P_2-Q)
\bigg]
\bigg\}.
\label{Eq:eff-vertex}
\eea
Plugging Eq.~\eqref{Eq:eff-vertex} in Eq.~\eqref{Eq:dkZk-LO}, one obtains [$E=(i\omega,{\bf 0})$]
\bea
\label{Eq:Zk-flow}
\partial_k Z_k (i\omega+\mu_\xi) &=& -\frac{Z_k}{2}\int_P \partial_k \mathcal{R}_k(P) \bigg[G_k^2 \tr\big[\hat D_{R,k}(P)\, \hat\p\cdot \bm{\sigma}\, \hat D_{R,k}(P)\big] \int_Q \tr\big[\hat D_{R,k}(Q)\big]\big(d(E-P+Q)+d(E+P-Q)\big)
\nonumber\\
&&\quad +2(H_k^2 + G_k H_k)\int_Q \tr\big[\hat D_{R,k}(P)\, \hat\p\cdot \bm{\sigma}\, \hat D_{R,k}(P)\hat D_{R,k}(Q) \big] \big(d(E-P+Q)+d(E+P-Q)\big)
  \bigg],
\eea
with the propagators $\hat D_{R,k}(P)$ and $d(P)$ given in Eq.~\eqref{Eq:reg_pert_prop}.
\end{widetext}
Calculating the integrals by expanding the integrand to linear order in $i\omega + \mu_\xi$, one obtains for $k\ll \mu$
\be
\label{Eq:Zk-flow_final}
k\partial_k Z_k = -\frac{1}{2} Z_k \bigg(\frac{\mu^2}{2\pi^2}\bigg)^2 \big(G_k^2 + H_k^2 + G_k H_k\big) + \mathcal{O}(k^0),
\ee
as shown in Appendix~\ref{app:int}.

Finally, we make a remark on the absence of the wave function renormalization for the $\psi$ field, $Z_{k,\psi}$. Following the procedure outlined above, one can derive an evolution equation for $Z_{k,\psi}$, and its structure is very similar to that of (\ref{Eq:Zk-flow}). The crucial difference between the two is that in the former all terms are proportional to $J_k^2$. That is to say, if one takes into account the scale dependence of $Z_{k,\psi}$ in the flow equation of $J_k$ itself, one concludes that the scale derivative of $J_k$ is still proportional to $J_k^2$ (as without the effect of $Z_{k,\psi}$), thus $J_k=0$ is a fixed point. We conclude that for the forthcoming fixed point analysis one is allowed to choose $J_k \equiv 0$, which yields $Z_{k,\psi}\equiv 1$.

\subsection{Fixed points \label{SS:FP}}
In terms of the dimensionless quantities $\bar G_k = \frac{\mu^2}{2\pi^2} G_k$ and $\bar H_k = \frac{\mu^2}{2\pi^2} H_k$, the nontrivial flow equations in the deep IR ($k\rightarrow 0$) limit read
\begin{subequations} \label{Eq:flow_eq_GHZ}
\bea
k\partial_k\big(Z_k \bar G_k\big) &=& -Z_k \bar G_k^2,\\
k\partial_k\big(Z_k \bar H_k\big) &=& \frac{1}{2}Z_k \bar G_k^2,\\
k\partial_k Z_k &=& -\frac{1}{2} Z_k \big(\bar G_k^2 + \bar H_k^2 + \bar G_k \bar H_k\big).
\eea
\end{subequations}
Here we remark that, although the situation with two independent couplings somewhat resembles that of an anisotropic interaction, the flow equations do not agree with those for the anisotropic Kondo model studied in Refs.~\cite{Solyom-Zawa,Kogan18,Kogan19}. As explained earlier around Eqs.~\eqref{Eq:isotrop-Kondo} and \eqref{Eq:identity}, our model does not break spin-rotational symmetry, as opposed to an anisotropic theory.
 
For the flow equations of the remaining couplings, we obtain
\begin{subequations}
\label{Eq:flowsLJD}
\bea
k\partial_k \bar{L}_k &=& \bar{L}_k \big(\bar{G}_k^2+\bar{H}_k^2+\bar{G}_k\bar{H}_k\big),\\
k\partial_k \bar{J}_k &=& \bar{J}_k^2,\\
k\partial_k \bar{\Delta}_k &=& \bar{\Delta}_k \bar{J}_k,
\eea
\end{subequations}
where we used the same rescaling as above, $\bar L_k = \frac{\mu^2}{2\pi^2} L_k$, $\bar J_k = \frac{\mu^2}{2\pi^2} J_k$,  and $\bar \Delta_k = \frac{\mu^2}{2\pi^2} \Delta_k$. We see, on the one hand, that the latter decouple from the system of flow equations for $\bar{G}_k$ and $\bar{H}_k$, and on the other hand that $\bar{L}_k=\bar{J}_k=\bar{\Delta}_k=0$ is an IR stable solution of (\ref{Eq:flowsLJD}) irrespective of the values for $\bar{G}_k$ and $\bar{H}_k$, provided that $\bar{J}_k>0$ and $\bar{\Delta}_k>0$.

\begin{figure}[!t]
    \centering
    \includegraphics[width=1\linewidth]{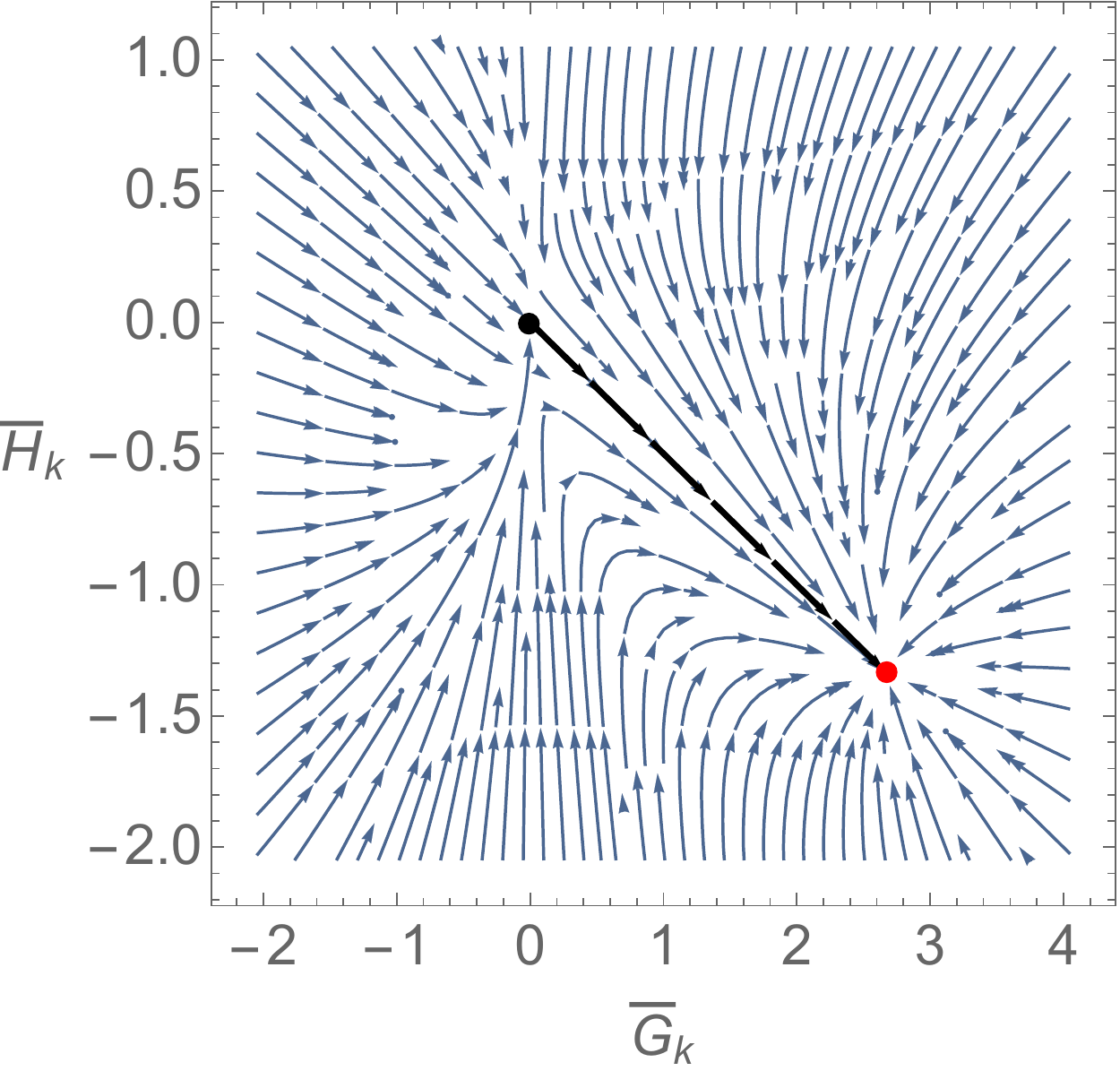}
    \caption{Phase portrait for $\bar{L}_k=\bar{J}_k=\bar{\Delta}_k=0$ showing the IR flow around the fixed points: the Gaussian one and the nontrivial fixed point with coordinates $(\bar G_\star, \bar H_\star)=(\frac{8}{3},-\frac{4}{3})$. \label{Fig:FP} }
\end{figure}

We now explore the fixed points of the flow equations~\eqref{Eq:flow_eq_GHZ} for $\bar{L}_k=\bar{J}_k=\bar{\Delta}_k=0$. Exploiting the third equation in the first two, one obtains
\begin{subequations}
\label{Eq:betaf}
\bea
k\partial_k \bar G_k &=& \frac{\bar G_k}{2} \big[-2 \bar G_k + \bar G_k^2 + \bar H_k^2 + \bar G_k \bar H_k\big],\\
k\partial_k \bar H_k &=& \frac{1}{2} \big[\bar G_k^2 + \bar H_k\big(\bar G_k^2 + \bar H_k^2 + \bar G_k \bar H_k\big)\big].
\eea
\end{subequations}
The fixed points are
\be
\bar G_\star = \bar H_\star = 0 \qquad \textrm{and}\qquad \bar G_\star = -2 \bar H_\star = \frac{8}{3}.
\ee
The stability matrix for the nontrivial fixed point is $(8/3)\mathbb{1}$, meaning that this fixed point is infrared ($k\to 0$) stable, as can be seen in Fig.~\ref{Fig:FP}.

Notice that the relations $\bar G_\star = -2 \bar H_\star \equiv 2\bar {\mathcal J}_\star$ lead to a spin-spin isotropic interaction. Indeed, using \eqref{Eq:identity}, the $G$- and $H$-type interactions combine in the fixed point as follows:
\bea
\label{Eq:spinspin}
&&\bar G_\star \big(\psi^\dagger\xi\big) \big(\psi^{\rm T}\xi^*\big) + \bar H_\star \big(\psi^\dagger\psi\big) \big(\xi^\dagger\xi\big)= \nonumber\\
&&\hspace{1cm}= 2\bar {\mathcal J}_\star \psi^*_i \xi_i \psi_j \xi^*_j - \bar {\mathcal J}_\star \psi^*_i \psi_i \xi^*_j \xi_j \nonumber\\
&&\hspace{1cm}=2 \bar {\mathcal J}_\star  \psi^*_i \xi_l \psi_j \xi_k^*\delta_{il}\delta_{jk}  - \bar {\mathcal J}_\star \psi^*_i \xi_l \psi_j \xi^*_k  \delta_{ij} \delta_{kl} \qquad \nonumber\\
&&\hspace{1cm}=\bar {\mathcal J}_\star \psi^*_i \xi_l  \psi_j \xi_k^* \vec{\sigma}_{ij} \cdot \vec{\sigma}_{kl}\equiv \bar {\mathcal J}_\star \vec{S}_{\psi} \cdot \vec{S}_{\xi},
\eea
where $\vec{S}_{\psi}=\psi^*_i \vec{\sigma}_{ij} \psi_j$,  $\vec{S}_{\xi}=\xi^*_i \vec{\sigma}_{ij} \xi_j$.

In the nontrivial fixed point $\bar {\mathcal J}_\star>0$. This corresponds to an antiferromagnetic interaction, which is essential for the Kondo effect.\footnote{Neglecting the $H$ coupling, i.e., setting $H_k\equiv 0$ (as done in Ref.~\cite{Kanazawa:2016ihl}), the nontrivial fixed point obtained with our approach is $\bar G_\star=2$.} Equation \eqref{Eq:spinspin} shows that, no matter what the microscopic interactions between the fields are, fluctuations drive the system toward an effective description that is spin-spin isotropic at the IR fixed point. This does not mean that throughout the RG flows the interaction is of the form dictated by this fixed point. On the contrary, even starting from a spin-spin isotropic theory at the UV scale, the non-perturbative RG flows do separate the $G$- and $H$-type interactions, just to combine them once again in the deep IR into a spin-spin isotropic form.

\bigskip
A few reflections on the nontrivial fixed point obtained are in order. Although the method applied, being based on an ansatz for the effective action appropriately regulated in the infrared, does not involve a small parameter, the nontrivial fixed point it gives seems to correspond to the one obtained in the perturbative RG from the weak-coupling $\beta$ function calculated to $\mathcal{O}(\mathcal{J}^3)$. In the perturbative context such a fixed point could have been questioned based on the fact that the value of the coupling at the fixed point invalidates the perturbative expansion. However, the fact that the current method yields a similar nontrivial fixed point might indicate that its existence might not be an artifact of the perturbative RG, after all. We wish to point out, however, that even though our approach does not deal with a small parameter, it does not necessarily guarantee that the obtained results are completely reliable in the strongly coupled region. One aspect is that, since in our method the exact four-point function between the impurity and the itinerant fermion appears in the flow equation of the wave function renormalization factor of the impurity field, one inevitably needs to use an input for it, and the result on the flow trajectories in the deep IR could depend on the approximation used. In particular, the absence of the strong coupling fixed point in our approach, corresponding to the exact screening and found with the numerical renormalization group method, could be the consequence of the one-loop approximation used for the four-point vertex. It would be interesting to investigate how a more sophisticated approximation of the four-point function, e.g., based on ladder resummation or using a $\Phi$-derivable approximation, would affect the flow equations determined with our method.

As a final note we mention that, had we added an internal index to the fermion with dimension $N_f$ in our ansatz~\eqref{Eq:Ansatz}, the position of the nontrivial fixed point would have been dumped with $1/N_f$. In the context of the Kondo effect, this $N_f$ is identified with the number of channels. For large enough $N_f$, it, therefore, becomes perturbative around the Gaussian fixed point and trustworthy even in perturbation theory. Such a behavior in the large $N_f$ limit is consistent with the CFT analysis~\cite{Kimura:2016zyv}. We also expect that in this context our present approach is more reliable.

\section{Summary and Outlook \label{sec:sum}}

In this paper we considered a field theoretical effective model to describe the scale evolution of the interaction between a Weyl fermion and a heavy magnetic impurity. Being interested in going beyond perturbation theory, we applied to this end the functional renormalization group method. The flow equations for the couplings and the wave function renormalization factor of the impurity field were obtained with the use of an inhomogeneous background field and an appropriate infrared regulator that cuts off fluctuations in a symmetric fashion around the Fermi surface of the itinerant fermion. Contrary to Anderson’s perturbative scaling approach, which preserves the spin-spin isotropic nature of the interaction, described by a single coupling, we found that the FRG formalism requires the independent evolution of two couplings between the itinerant fermion and the impurity, both respecting the global $SU(2)$ symmetry of the system. 

By investigating the flow equations in the momentum limit relevant for the Kondo effect, we determined the fixed point structure of the model for vanishing pairing gap. Beside the Gaussian fixed point, we have found also an interacting, infrared stable fixed point with vanishing self-coupling of the itinerant fermion and the impurity. For a general initial condition the system explores during its scale evolution regions of the two dimensional coupling space in which the interaction with the impurity is not spin-spin isotropic. The isotropic nature of the interaction can be preserved in the deep IR on a line that connects in the space of the couplings the Gaussian fixed point to the nontrivial one. On this line the evolution is governed by a single $\beta$ function, as in the case of the poor man's renormalization group procedure. As mentioned in the previous section, the interacting, infrared stable fixed point was already seen in perturbation theory, where its actual existence seemed questionable. We argued that the appearance of this fixed point might not be an artifact of perturbation theory, after all, but also stressed that, due to the approximation used for the four-point vertex, the reliability of our results in the strongly coupled regime is not guaranteed. One should also note that although the intermediate-coupling fixed point does not determine the low-energy behavior of the Kondo model with exact screening, where it is known that the coupling flows to infinity, it can be relevant in other fermionic systems.

We would also like to emphasize that the FRG formalism seems to be a unique analytic approach that can reveal a nontrivial interacting IR fixed point of a quantum impurity system at the non-perturbative level. Compared with the CFT analysis, which is another powerful non-perturbative approach to this problem, this is a significant advantage. In fact, one cannot say anything {\it a priori} on the existence of such a nontrivial fixed point in the framework of the CFT method itself, and hence its existence is always an assumption. From this point of view, it is quite important to establish nontrivial IR fixed points based on an alternative non-perturbative method as discussed in this paper.

There are several future directions in which the presented application of the FRG flow equations could be of interest.
First of all, it would be important to compute physical observables based on our formalism. In particular, the thermodynamic quantities, e.g., specific heat and magnetic susceptibility, are primary observables computed in various models. It would be desirable to compute these quantities in the FRG formalism and compare them with known results and experimental measurements. In addition to the thermodynamic quantities, it would also be meaningful to compute transport coefficients, in particular, the electric conductivity. One is also interested in going beyond the approximation scheme presented here, e.g., taking into account the momentum dependence of the wave function renormalization factors and/or the four-point vertex entering their flow equations. Inclusion of higher order interactions can also be considered an important direction for a forthcoming work.

Generalization of the Kondo effect beyond the unitary group interaction, e.g., orthogonal group \cite{Beri:2012tr,Mitchell--Liberman} and symplectic group interactions~\cite{Stadler--Mitchell16, Li--Konig--Vayrynen},  has also attracted attention; see also Ref.~\cite{Kimura21}. From this point of view, it would be interesting to generalize the FRG analysis presented here to real Majorana fermions in order to discuss the Kondo problem in aforementioned systems.

\vspace{-0.08cm}
\section*{Acknowledgements} G.~F. and Zs.~Sz. very much appreciate the hospitality of the University of Burgundy, where this research project was initiated as a collaboration within the framework of France-Romania-Hungary Research Network, whose support is kindly acknowledged. G.~F. was also supported from the Hungarian National Research, Development and Innovation Fund (NRDI Fund) under Project No. FK142594.
The work of T.~K. was supported by EIPHI Graduate School (No.~ANR-17-EURE-0002) and Bourgogne-Franche-Comt{\'e} region.

\appendix
\section{Matrix elements of $\Sigma^{ab}_k[\Phi](P,-Q)$ \label{app:Sigma_elem}}

With the notations $(A \star B)(K) := \int_R A(R) B(K-R)$ and $(A \bullet B)(K) := \int_R A(R) B(K+R)$, the elements of the $4\times4$ block matrix $\Sigma^{ab}_k[\Phi](P,-Q)\equiv\Sigma^{ab}$ defined below \eqref{Eq:Dyson} are
\bea
  \label{Eq:Sigma_elements}
    \Sigma^{ab}_{11}&=&\Delta_k \sigma_2^{ab}\,\deltabar(Q-P)-2 J_k (\psi_a^* \star \psi_b^*)(Q-P),  \nonumber\\
    \Sigma^{ab}_{22}&=&\Delta_k \sigma_2^{ab}\,\deltabar(P-Q) -2 J_k (\psi_a \star \psi_b)(P-Q),\nonumber\\
    \Sigma^{ab}_{12}&=&-G_k (\xi_a^*\bullet \xi_b)(P-Q) - \delta_{a b} H_k (\xi_s^* \bullet \xi_s)(P-Q) \nonumber\\
    &-&2 J_k \Big[\delta_{ab} (\psi_s^* \bullet \psi_s)(P-Q) - (\psi_a^*\bullet \psi_b)(P-Q)\Big],\nonumber\\
    \Sigma^{ab}_{21}&=&G_k (\xi_b^*\bullet \xi_a)(P-Q) + \delta_{a b} H_k (\xi_s^* \bullet \xi_s)(P-Q)\nonumber\\
    &+&2 J_k \Big[\delta_{ab} (\psi_s^* \bullet \psi_s)(P-Q) - (\psi_b^*\bullet \psi_a)(P-Q)\Big], \nonumber\\
    \Sigma^{ab}_{13}&=&G_k(\xi^*_a \star \psi^*_b)(Q-P) - H_k(\psi^*_a \star \xi^*_b)(Q-P), \nonumber\\
    \Sigma^{ab}_{14}&=&\delta_{ab}G_k(\psi^*_s \bullet \xi_s)(P-Q) + H_k (\psi^*_a \bullet \xi_b)(P-Q),\nonumber\\
    \Sigma^{ab}_{23}&=&\delta_{ab}G_k( \psi_s \bullet \xi^*_s )(Q-P) + H_k(\psi_a \bullet \xi^*_b)(Q-P), \nonumber\\
    \Sigma^{ab}_{24}&=&G_k(\xi_a \star \psi_b)(P-Q) - H_k (\psi_a \star \xi_b)(P-Q),\nonumber\\
    \Sigma^{ab}_{33}&=&-2 L_k(\xi^*_a \star \xi^*_b)(Q-P), \nonumber\\
    \Sigma^{ab}_{44}&=& -2 L_k(\xi_a \star \xi_b)(P-Q), \nonumber\\
    \Sigma^{ab}_{34}&=&-G_k(\psi^*_a \bullet \psi_b)(P-Q) -\delta_{ab} H_k(\psi^*_s \bullet \psi_s)(P-Q)\nonumber\\
     &-& 2 L_k \Big[\delta_{ab} (\xi_s^*\bullet \xi_s)(P-Q) - (\xi_a^* \bullet \xi_b)(P-Q)\Big]\!, \nonumber \\
\eea
The matrix elements that are not listed can be generated using the property $\Sigma^{ba}_k[\Phi](-Q,P)=-\Sigma^{ab}_k[\Phi](P,-Q)$, as exemplified above for the 21 element.

\section{Complete set of flow equations for the couplings and the evaluation of integrals \label{app:int}}

The complete set of flow equations for the couplings coming from the $n=2$ term of (\ref{Eq:WE_field_exp}) can be given in terms of nine integrals $I_i$, $i=1,\dots, 9$:
\begin{subequations}
\bea
Z_k^{-1}\partial_k (Z_k G_k) &=& -2 G_k^2 I_1- 2 G_k H_k(I_1+I_2)\nonumber\\
&& - 2 G_k (J_k I_3 + L_k I_4),\\
Z_k^{-1}\partial_k (Z_k H_k) &=&  -G_k^2 I_2 - H_k^2(I_1+I_2) \nonumber\\
&&+2 G_k (J_k I_5 + L_k I_4) \nonumber\\
&&+ 2 H_k(J_k I_9 + L_kI_4),\\
\partial_k J_k &=& -2 J_k^2(I_6+I_7)\nonumber\\
&&-\Big(\frac{1}{2}G_k^2-H_k^2-G_kH_k\Big)I_4,\\
Z_k^{-2}\partial_k \big(Z_k^2 L_k\big) &=& -\frac{1}{2}G_k^2 I_6 - 2 L_k^2(I_4+I_8) \nonumber\\
&&+ H_k(G_k+H_k) I_9,\\
\partial_k \Delta_k &=& -2\Delta_k J_k I_7,
\eea
\end{subequations}
where for the sake of completeness we indicated the leading order flow equation of $\Delta_k$ obtained by expanding in $\Delta_k$. However, as explained in the main text, we work in the $\Delta_k=0$ fixed point of that equation. Although the integrals introduced above are not linearly independent (e.g. $I_5=2I_3-I_6$), we prefer using them as they compactify the flow equations. The decomposition of the integrals into elementary ones, as well as their explicit expression is as follows:
\begin{subequations}
\bea
I_3&=&\frac{1}{6}\int_P\tilde\partial_k \Big[2\big(\tr \hat D_{R,k}(P)\big)^2-\tr\hat D^2_{R,k}(P)\Big] \nonumber\\
&=&\int_P\tilde\partial_k \frac{(i p_0 + \mu)^2 - \frac{p_R^2}{3}}{\big[(i p_0 + \mu)^2 - p_R^2\big]^2}, \\
I_5&=&\frac{1}{6}\int_P\tilde\partial_k \Big[\big(\tr \hat D_{R,k}(P)\big)^2+\tr\hat D^2_{R,k}(P)\Big]\nonumber\\
&=&\int_P\tilde\partial_k \frac{(i p_0 + \mu)^2 + \frac{p_R^2}{3}}{\big[(i p_0 + \mu)^2 - p_R^2\big]^2}, \\ 
I_6&=&\frac{1}{2}\int_P\tilde\partial_k\Big[\big(\tr \hat D_{R,k}(P)\big)^2-\tr\hat D^2_{R,k}(P)\Big] \nonumber\\
&=&\int_P\tilde\partial_k\frac{1}{(i p_0 + \mu)^2 - p_R^2},\\
I_7&=& \frac{1}{2}\int_P\tilde\partial_k\Big[\tr\hat D_{R,k}(P)\tr\hat D_{R,k}(-P)\nonumber\\
&&\hspace{1.2cm}-\tr\big(\hat D_{R,k}(P)\hat D_{R,k}(-P)\big)\Big] \nonumber\\
&=& \frac{1}{2}\int_P\tilde\partial_k \tr\Big[\hat D_{R,k}^{\rm T}(-P)\sigma_2\hat D_{R,k}(P)\sigma_2 \Big]\nonumber\\
&=& \int_P\tilde\partial_k\frac{(\mu+i p_0)(\mu-i p_0)+p_R^2}{\big[(i p_0 +\mu)^2 -p_R^2\big]\big[(i p_0 -\mu)^2 - p_R^2\big]}, \nonumber\\ \\
I_9 &=&  \frac{1}{2}\int_P\tilde\partial_k \tr \hat D_{R,k}^2(P) \nonumber\\
&=& \int_P\tilde\partial_k \frac{(i p_0 + \mu)^2 + p_R^2}{\big[(i p_0 + \mu)^2 - p_R^2\big]^2},\\
I_4&=&\frac{1}{2}\int_P\tilde\partial_k\Big[\big(\tr \hat d(P)\big)^2-\tr\hat d^2(P)\Big] \nonumber\\
&=& \int_P\tilde\partial_k \frac{1}{(i p_0+\mu_\xi)^2}\equiv 0,\\
I_8&=&\frac{1}{2}\int_P\tilde\partial_k\Big[\tr\hat d(P)\tr\hat d(-P)-\tr\big(\hat d(P)\hat d(-P)\big)\Big]  \nonumber\\
&=& \int_P\tilde\partial_k \frac{1}{(i p_0+\mu_\xi)(\mu_\xi - i p_0)}\equiv 0.
\eea
\end{subequations}
Integrals $I_1$ and $I_2$ are defined below \eqref{Eq:G_H_flows}. The second form of $I_7$ appears in the flow equation of $\Delta_k$. The above integrals can be easily evaluated with the regulator given in \eqref{Eq:reg_fv}. It turns out that with the exception of $I_1, I_2$ and $I_7$ their contribution is regular for $k\to 0$. In fact, $I_7=-\mu^2/(4\pi^2 k)$ for small $k$, but since the flow equations of $\bar J_k,\bar \Delta_k$, and $\bar L_k$ do not influence those of $\bar G_k$ and $\bar H_k$, only the integrals $I_1$ and $I_2$ prove relevant for the Kondo effect.

As an example we calculate the integral $I_1$ for $\mu>0$ and $\mu_\xi<0$. After a decomposition in partial fractions one obtains from \eqref{Eq:I1}
\be
I_1(Q)=\frac{1}{2}\sum_{\pm}\int_P \partial_k \mathcal{R}_k\frac{1}{[i(p_0+\omega)+\mu \pm p_R]^2(i p_0 + \mu_\xi)}. 
\ee
Only the term with the minus sign gives contribution to the Kondo effect, as the other term gives vanishing contribution in the $k\to 0$ limit. Contour integration results in
\bea
&&\hspace{-1cm}\int_{-\infty}^\infty \frac{{\rm d} p_0}{2\pi} \frac{1}{[i(p_0+\omega)+\mu \pm p_R]^2(i p_0 + \mu_\xi)} \nonumber\\
&&\hspace{1cm}= -\frac{\Theta(\mu-p_R)}{(i\omega-\mu_\xi + \mu \pm p_R)^2}.
\eea
Analytic continuation to a vanishing real frequency means taking the limit $i\omega -\mu_\xi\to 0$. Using the identity $\Theta(\mu-p_R) \partial_k \mathcal{R}_k(P)=-\Theta(\mu-k\le p \le \mu)$, a simple integration over $p$ yields in this limit the result given in Eq.~\eqref{Eq:I1_2_res}.

Another integral of interest appears in the flow equation for $Z_k$. After doing the traces in \eqref{Eq:Zk-flow}, namely, 
\begin{subequations}
\bea
 &&\hspace{-0.5cm}\tr\big[\hat D_{R,k}(P)\, \hat\p\cdot \bm{\sigma}\, \hat D_{R,k}(P)\big]\nonumber\\
 &&=4 p_R\frac{-i p_0 - \mu}{\big[(i p_0+\mu)^2- p^2_R\big]^2},\\
 &&\hspace{-0.5cm}\tr\big[\hat D_{R,k}(P)\, \hat\p\cdot \bm{\sigma}\, \hat D_{R,k}(P)\hat D_{R,k}(Q) \big]\nonumber\\
 &&=4 p_R\frac{(i p_0 + \mu)(i q_0 + \mu)}{\big[(i p_0+\mu)^2- p^2_R\big]^2 \big[(i q_0+\mu)^2- q^2_R\big]}, \qquad
\eea
\end{subequations}
\begin{widetext}
\hspace{-0.35cm}one notices that we need to evaluate the following double integral [$E=(i\omega,0)$]
\be
I_{10}(E)=\int_P\int_Q p_R\partial_k\mathcal{R}_k(P)\frac{(i p_0 + \mu)(i q_0 + \mu)}{\big[(i p_0+\mu)^2-p^2_R\big]^2 \big[(i q_0+\mu)^2-q^2_R\big]}\sum_{\pm}\frac{1}{-i(\omega \mp p_0 \pm q_0) - \mu_\xi}.
\ee
The $q_0$ and $p_0$ integrals can be dealt with contour integration. The $q_0$ integral results in
\begin{subequations}
\bea
A(p_0;p_R,q_R)&:=&\int_{-\infty}^\infty\frac{{\rm d} q_0}{2\pi}\frac{i q_0 + \mu}{\big[(i q_0+\mu)^2-q^2_R\big]}\frac{1}{-i(\omega - p_0 + q_0) - \mu_\xi}\nonumber\\
&=&-\frac{1}{2}\bigg[\frac{1}{i(\omega-p_0)-\mu+\mu_\xi-q_R} + \frac{\Theta(\mu-q_R)}{i(\omega-p_0)-\mu+\mu_\xi+q_R}\bigg],
\\
B(p_0;p_R,q_R)&:=&\int_{-\infty}^\infty\frac{{\rm d} q_0}{2\pi}\frac{i q_0 + \mu}{\big[(i q_0+\mu)^2-q^2_R\big]}\frac{1}{-i(\omega + p_0 - q_0) - \mu_\xi}=
\frac{1}{2}\,\frac{\Theta(q_R-\mu)}{i(p_0+\omega)+\mu+\mu_\xi-q_R}.
\eea
\end{subequations}
which upon integration over $p_0$ gives
\begin{subequations}
\bea
\int_{-\infty}^\infty\frac{{\rm d} p_0}{2\pi}\frac{i p_0 + \mu}{\big[(i p_0+\mu)^2-p^2_R\big]^2} A(p_0;p_R,q_R) =
\frac{\Theta(p_R-\mu)}{8 p_R} \bigg[\frac{1}{(i\omega+\mu_\xi-p_R-q_R)^2} + \frac{\Theta(\mu-p_R)}{(i\omega + \mu_\xi - q_R +q_R)^2}\bigg]\,, \\
\int_{-\infty}^\infty\frac{{\rm d} p_0}{2\pi}\frac{i p_0 + \mu}{\big[(i p_0+\mu)^2-p^2_R\big]^2} B(p_0;p_R,q_R) = \frac{\Theta(q_R-\mu)}{8 p_R}
\bigg[\frac{1}{(i\omega+\mu_\xi-p_R-q_R)^2} - \frac{\Theta(\mu-p_R)}{(i\omega + \mu_\xi + p_R -q_R)^2}\bigg].
\eea
\end{subequations}
Expanding the above expressions to linear order in $i\omega + \mu_\xi$, one easily evaluates the remaining integrals over $\p$ and $\q$, which decouple due to the regulator, to obtain the result given in \eqref{Eq:Zk-flow_final}.

Finally, we sketch the calculation of the integrals $I_1^D$ and $I_2^D$ introduced in \eqref{Eq:pert_I1_I2} in relation to the perturbative scaling. Contour integration gives
\begin{subequations}
\bea
\int_{-\infty}^\infty\frac{{\rm d} p_0}{2\pi} \frac{i(p_0+\omega) + \mu}{\big(i(p_0+\omega)+\mu\big)^2-p^2}\,\frac{1}{i p_0+\mu_\xi} 
&=& \frac{\Theta(\mu - p) (\mu + i\omega -\mu_\xi)}{p^2 - (\mu + i\omega - \mu_\xi)^2} - \frac{1}{2} \frac{\Theta(p-\mu)}{\mu + p + i\omega - \mu_\xi}\,, \qquad
\\
\int_{-\infty}^\infty\frac{{\rm d} p_0}{2\pi} \frac{\mu-i(p_0+\omega)}{\big(i(p_0+\omega)-\mu\big)^2-p^2}\, \frac{1}{i p_0+\mu_\xi}
&=& \frac{1}{2} \frac{\Theta(p-\mu)}{p - \mu + i\omega - \mu_\xi}\,.
\eea
\end{subequations}
After taking the limit $i\omega -\mu_\xi\to 0$, one observes that there is no logarithmic contribution to $I_1^D$ for $p>\mu+D$, region in which one should apply also an ultraviolet cutoff $\Lambda$. Then, the leading order result for $D\ll \mu$ is
\begin{subequations}
\bea
\lim_{i\omega-\mu_\xi\to 0}\!I^D_1(-i\omega,\bm{0}) &=& \frac{\mu}{2\pi^2} \int_0^{\mu-D} {\rm d} p \frac{p^2}{p^2-\mu^2} \approx \frac{\mu^2}{4\pi^2}\ln\frac{D}{\mu},
\\
\lim_{i\omega-\mu_\xi\to 0}\!I^D_2(-i\omega,\bm{0}) &=& \frac{1}{4\pi^2} \int_{\mu+D}^\Lambda {\rm d} p \frac{p^2}{p-\mu} \approx -\frac{\mu^2}{4\pi^2}\ln\frac{D}{\mu},
\eea
\end{subequations}
as stated in \eqref{Eq:pert_I1_I2_sing}.
\end{widetext}

\end{document}